\documentclass[useAMS,usenatbib]{mn2e}
\usepackage{graphics}
\usepackage{graphicx}
\usepackage{epstopdf}
\usepackage[dvips]{epsfig}

\newcommand{\uv}{\mbox{$u$-$v$}}

\title[Characterisation of Long Baseline Calibrators]{Characterisation of Long Baseline Calibrators at 2.3~GHz}
\author[F. Hungwe et. al]{F. Hungwe$^{1,2}$\thanks{E-mail:
faith@hartrao.ac.za (FH); roopesh.ojha@gmail.com (RO)}, R. Ojha$^{3,4}$, R.S Booth$^{1,2}$, M.F Bietenholz$^{1,5}$, A. Collioud$^{6,7}$, P. Charlot$^{6,7}$
\newauthor D. Boboltz$^{4}$, A.L Fey$^{4}$\\
$^{1}$ Dept. of Physics and Electronics, Rhodes University, P.O Box 94, Grahamstown, 6140, South Africa\\
$^{2}$ Hartebeesthoek Radio Astronomy Observatory, P.O Box 443, Krugersdorp, 1740, South Africa\\
$^{3}$ NVI, Inc., 7257D Hanover Parkway, Greenbelt, MD 20770, USA\\
$^{4}$ United States Naval Observatory, 3450 Massachusetts Ave., NW, Washington D.C 20392, USA\\
$^{5}$ Dept. of Physics and Astronomy, York University, Toronto, M3J 1P3, Ontario, Canada\\
$^{6}$ Universit\'e de Bordeaux, Observatoire Aquitain des Sciences de l'Univers, BP 89, 33271 Floirac Cedex, France\\
$^{7}$ CNRS, Laboratoire d'Astrophysique de Bordeaux-UMR 5804, BP 89, 33271 Floirac Cedex, France
}

\begin{document}

\date{Accepted 2011 June 11. Received 2011 May 24; in original form 2010 October 29}

\pagerange{\pageref{firstpage}--\pageref{lastpage}} 
\pubyear{2010}

\maketitle

\label{firstpage}

\begin{abstract}
We present a detailed multi-epoch analysis of 31 potential southern hemisphere radio calibrators that were originally observed as part of a program to maintain the International Celestial Reference Frame (ICRF). At radio wavelengths, the primary calibrators are Active Galactic Nuclei (AGN), powerful radio emitters which exist at the centre of most galaxies. These are known to vary at all wavelengths at which they have been observed. By determining the amount of radio source structure and variability of these AGN, we determine their suitability as phase calibrators for long baseline radio interferometry at 2.3~GHz. For this purpose, we have used a set of complementary metrics to classify these 31 southern sources into five categories pertaining to their suitability as VLBI calibrators.  We find that all of the sources in our sample would be good interferometric calibrators and almost ninety per cent would be very good calibrators.
\end{abstract}


\section[]{Introduction}
High angular resolution observations of weak radio sources (where self calibration is not possible)  require calibrator sources for correction of systematic effects and effects of the atmosphere on the measured visibilities. Atmospheric fluctuations cause perturbations in visibility phase which, if not corrected, seriously limit both the sensitivity and image quality of an interferometric array. Phase calibrators are also required for astrometric observations. An ideal calibrator would look the same on all observing baselines. It should be bright, unresolved, or at least compact and should not vary.  
A calibrator 
source should also be separated from the target sources by as small an angle as possible in order to look along the same line of sight through the atmosphere. Therefore, it is desirable to have calibrator sources evenly distributed across the whole sky. In practice, at radio wavelengths, calibrators are mostly Active Galactic Nuclei (AGN), whose fundamental source of power is believed to be the accretion of matter onto a super-massive black hole \citep{Rees:97}. They are known to vary at every wavelength at which they have been studied. AGN are very compact and isotropically distributed around the sky. They are also very distant objects and therefore, generally, have no discernible proper motions on the sky. It is these qualities that make them suitable as calibrators at radio wavelengths. 

Due to the limited number of radio telescopes, few surveys for calibrators have been carried out in the southern hemisphere \citep{Ojha:04a, Ojha:05, Fey:04a, Fey:04b, Fey:06}. Hence, there are fewer known calibrators in the south. A major expansion of radio astronomy observing capability is underway in the southern hemisphere. Two SKA (Square Kilometre Array)  precursors, the South African MeerKAT (Karoo Array Telescope, \citet{Booth:09}) and ASKAP (Australian SKA Pathfinder, \citet{Johnston:08}) are presently under construction, leading to the SKA itself. It is clear that interferometry and Very Long Baseline Interferometry (VLBI) in the southern hemisphere needs a dense network of calibration sources at high resolutions and a range of frequencies. MeerKAT is the South African SKA demonstrator telescope. When completed, it will be the largest radio telescope array in the southern hemisphere. It will operate in two frequency ranges, 0.58-2.5~GHz and 8-14.5~GHZ \citep{Booth:09}. MeerKAT will participate in VLBI observations with the European VLBI Network and other VLBI arrays. The SKA will operate in the frequency range 0.7 to 25~GHz \citep{Carilli:04,Schilizzi:07} and is expected to have much longer baselines for which a southern VLBI calibrator list will be essential.

The United States Naval Observatory, in collaboration with NASA, the Laboratoire d'Astrophysique de Bordeaux (LAB) and the National Radio Astronomy Observatory (NRAO\footnote{The National Radio Astronomy Observatory is a facility of the National Science Foundation operated under cooperative agreement by Associated Universities, Inc}) has, since 1994 been observing AGN, once every two months, through the Research and Development VLBI (RDV) experiments. This has led to a wealth of data of sources observed simultaneously at 2.3~GHz and 8.4~GHz at regular epochs. Images of these AGN form the USNO's Radio Reference Frame Image Database (RRFID) and LAB's Bordeaux VLBI Image Database (BVID)\footnote{The web site for the RRFID is located at http://rorf.usno.navy.mil/rrfid.shtml and the BVID at http://www.obs.u-bordeaux1.fr/BVID/} \citep{Fey:96, Fey:97,Fey:00,Colli:09}. The RRFID is a database of about 6700 images of over 700 AGN sources compiled from geodetic and astrometric VLBI experiments. The BVID contains over 1800 images of 824 radio sources. It is the goal of the ongoing RDV program to image the radio reference frame sources on a regular basis, to monitor them for variability and structural change. \citet{Piner:07} discuss jet kinematics in a subset of the RRFID sources, concentrating on the 8.4~GHz observations. Their survey is made up of RRFID sources that have been observed at 3 or more epochs from July 1994 to December 1998.

In this paper, we focus on 2.3 GHz observations of a sample of 31 potential southern VLBI calibrators (with declinations between 0 and $-60^{\circ}$) from the kinematic survey of \citet{Piner:07}. The choice of 2.3 GHz was made because both MeerKAT and ASKAP will be operating at frequencies close to this. We seek to characterise this selection of southern radio sources, and determine their suitability as calibrators for southern VLBI experiments, especially those using MeerKAT and the SKA when completed.

In the following sections, we will describe the observations and data reduction process. We will then describe the imaging and modelfitting process and go on to analyse the data and discuss the results.

\begin{table*}
\begin{center}
\begin{tabular}{llclcccccc}
\hline 
B1950  & Other&Optical &\phantom{00}$z$&\multicolumn{2}{l}{Co-ordinates}&\multicolumn{2}{c}{Latest Flux Densities}&Latest\\
Source Name&Name&ID&&RA (hh mm ss)&DEC (deg mm ss)&$S_{total}$& $S_{core}$&Epoch\\
\hline 
\hline
0003-066& &BL Lac&0.35$^{1}$&00 06 13.89288849&-06 23 35.3353162&2.29&1.79&Jan 2008\\
0104-408&&BL Lac&0.58$^{2}$&01 06 45.10796851&-40 34 19.9602291&1.51&1.52&Jan 2008\\
0238-084&NGC1052&Galaxy&0.005$^{3}$&02 41 04.79850256&-08 15 20.7517956&0.53&0.26&Dec 2007\\	
0336-019&CTA26&Quasar&0.85$^{4}$&03 39 30.93778751&-01 46 35.8041062&2.53&2.16&Jan 2008\\
0402-362&&Quasar&1.42$^{5}$&04 03 53.74989835&-36 05 01.9131085&0.91&0.99&Mar 2007\\
0454-234&&Quasar&1.00$^{1}$&04 57 03.17922863&-23 24 52.0201418&3.41&3.01&Dec 2007\\
0458-020&&Quasar&2.29$^{6}$&05 01 12.80988366&-01 59 14.2562534&0.82&0.65&Dec 2007\\
0727-115&&Quasar&1.59$^{7}$&07 30 19.11247420&-11 41 12.6005110&3.77&3.47&Jan 2008\\
0919-260&&Quasar&2.30$^{8}$&09 21 29.35385535&-26 18 43.3861684&1.58&1.50&Sep 2006\\
0920-397&&Quasar&0.59$^{9}$&09 22 46.41826064&-39 59 35.0683561&1.56&1.49&Jan 2008\\
1034-293&&Quasar&0.31$^{1}$&10 37 16.07973476&-29 34 02.8133345&1.60&1.51&Jan 2008\\
1124-186&&Quasar&1.05$^{10}$&11 27 04.39244958&-18 57 17.4416582&1.16&1.16&Jan 2008\\
1144-379&&Quasar&1.05$^{1}$&11 47 01.37070177&-38 12 11.0234199&1.09&1.09&Jan 2007\\
1145-071&&Quasar&1.34$^{11}$&11 47 51.55402876&-07 24 41.1410887&0.95&0.81&Jan 2007\\
1253-055&3C279&Quasar&0.54$^{12}$&12 56 11.16656541&-05 47 21.5247030&7.60&7.30&Dec 1998\\
1255-316&&Quasar&1.92$^{13}$&12 57 59.06081737&-31 55 16.8516980&1.51&1.38&Jan 2008\\
1313-333&&Quasar&1.21$^{14}$&13 16 07.98593995&-33 38 59.1725057&0.76&0.53&Feb 2004\\
1334-127&&Quasar&0.54$^{15}$&13 37 39.78277768&-12 57 24.6932620&2.61&2.57&Jan 2008\\
1351-018&&Quasar&3.71$^{16}$&13 54 06.89532213&-02 06 03.1904447&1.00&0.98&Jan 2008\\
1424-418&&Quasar&1.52$^{2}$&14 27 56.29756536&-42 06 19.4375991&2.04&1.66&Jan 2008\\
1451-375&&Quasar&0.31$^{17}$&14 54 27.40975442&-37 47 33.1448724&0.56&0.51&Jul 2006\\
1514-241&&BL Lac&0.05$^{18}$&15 17 41.81313221&-24 22 19.4760251&2.94&2.24&Jan 2007\\
1622-253&&Quasar&0.79$^{19}$&16 25 46.89164010&-25 27 38.3267989&1.01&0.99&Jan 2008\\
1741-038&&Quasar&1.05$^{2}$&17 43 58.85613396&-03 50 04.6166450&4.78&4.73&Jan 2008\\
1908-201&&Quasar&1.12$^{20}$&19 11 09.65289198&-20 06 55.1089891&2.10&1.78&Dec 2007\\
1921-293&&Quasar&0.35$^{9}$&19 24 51.05595514&-29 14 30.1210524&6.18&5.43&Dec 2007\\
1954-388&&Quasar&0.63$^{21}$&19 57 59.81927470&-38 45 06.3557585&2.75&2.69&Jan 2008\\
1958-179&&Quasar&0.65$^{21}$&20 00 57.09044485&-17 48 57.6725440&1.53&1.56&Dec 2007\\
2052-474&&Quasar&1.49$^{13}$&20 56 16.35981874&-47 14 47.6276461&1.37&1.43&Dec 2007\\
2243-123&&Quasar&0.63$^{21}$&22 46 18.23197613&-12 06 51.2774796&2.57&1.33&Jan 2008\\
2255-282&&Quasar&0.93$^{5}$&22 58 05.96288481&-27 58 21.2567425&0.95&0.87&Jan 2008\\	    

\hline
\end{tabular}
\end{center}
\caption{Optical and radio properties of the sample. The positions shown above are the most recent ICRF positions  \citep{Fey:09}. $S_{core}$ is the flux density in the Gaussian component fitted to the core of the latest epoch, 
while the total flux density $S_{total}$ is the sum of the CLEAN components.
$^1$\citet{Stickel:89};
$^2$\citet{White:88};
$^3$\citet{Deni:05};
$^4$\citet{Wills:78};
$^5$\citet{Pete:76};
$^6$\citet{Stritt:74};
$^7$\citet{Zensus:02};
$^8$\citet{Wright:79};
$^9$\citet{Hewitt:89};
$^{10}$\citet{Linfield:89};
$^{11}$\citet{Wilkes:86};
$^{12}$\citet{Marz:96};
$^{13}$\citet{Jauncey:84};
$^{14}$\citet{Jauncey:82};
$^{15}$\citet{Stickel:93};
$^{16}$\citet{Osmer:94};
$^{17}$\citet{Jones:04};
$^{18}$\citet{Jones:09};
$^{19}$\citet{di:94};
$^{20}$\citet{Halpern:03};
$^{21}$\citet{Browne:75}.}
\label{tsouthernsample}
\end{table*}

\section[]{Observations and Data Reduction}
\begin{figure}
 \centering
 \includegraphics[height=6cm]{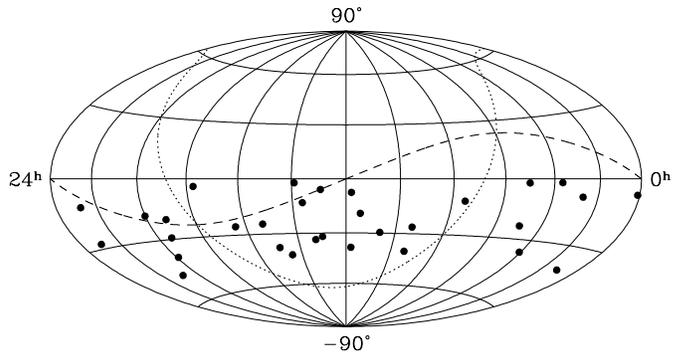}
  \vspace{-20pt}
  \caption{Sky distribution of the southern sources plotted on an Aitoff equal-area projection of the celestial sphere. The dotted line represents the Galactic plane while the dashed line is the ecliptic.}
  \vspace{-10pt}
 \label{skycoverage}
 \end{figure}
 
 \begin{figure*}
 \centering
 \includegraphics[height=7.6cm]{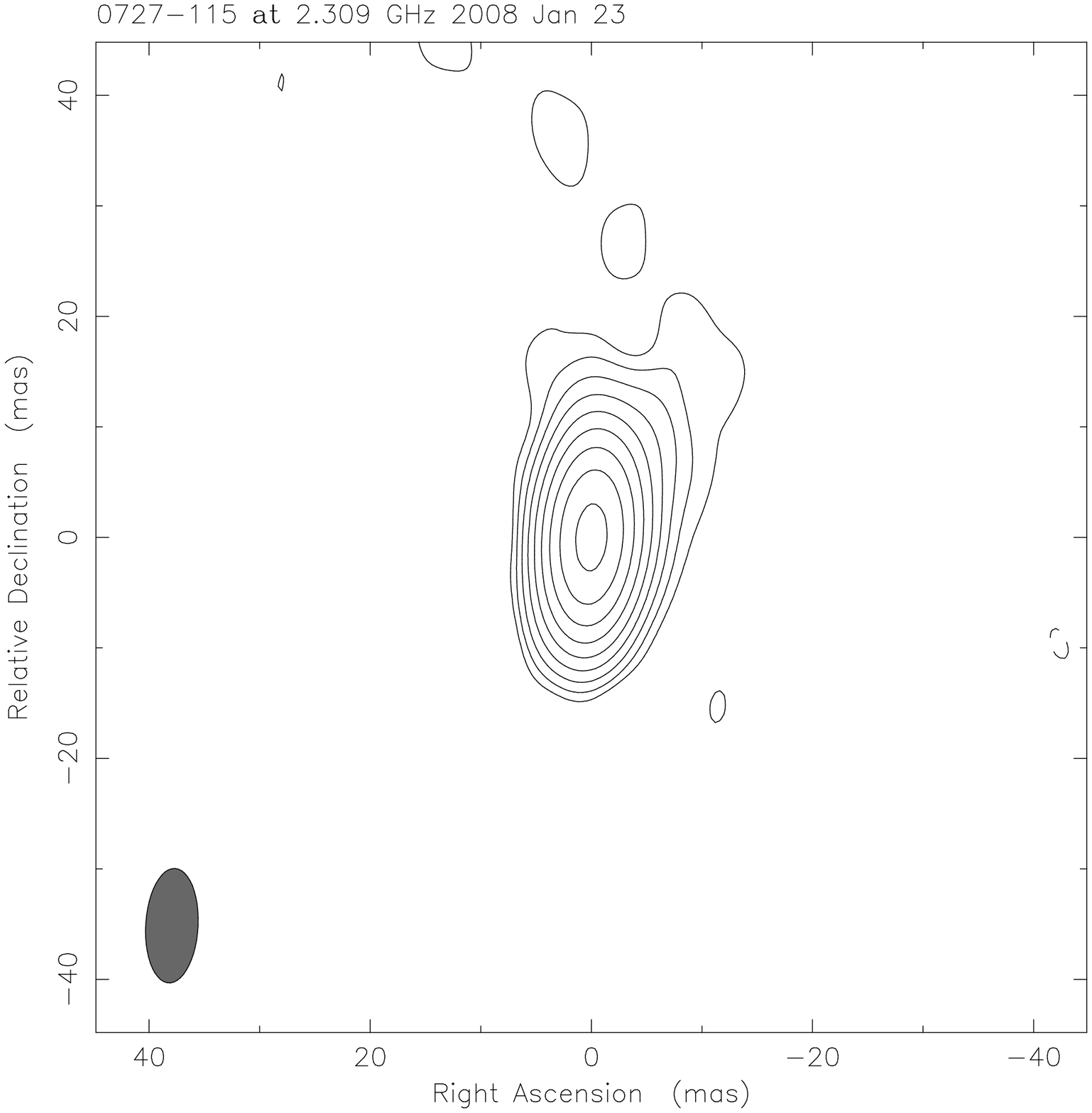}
 \includegraphics[height=7.6cm]{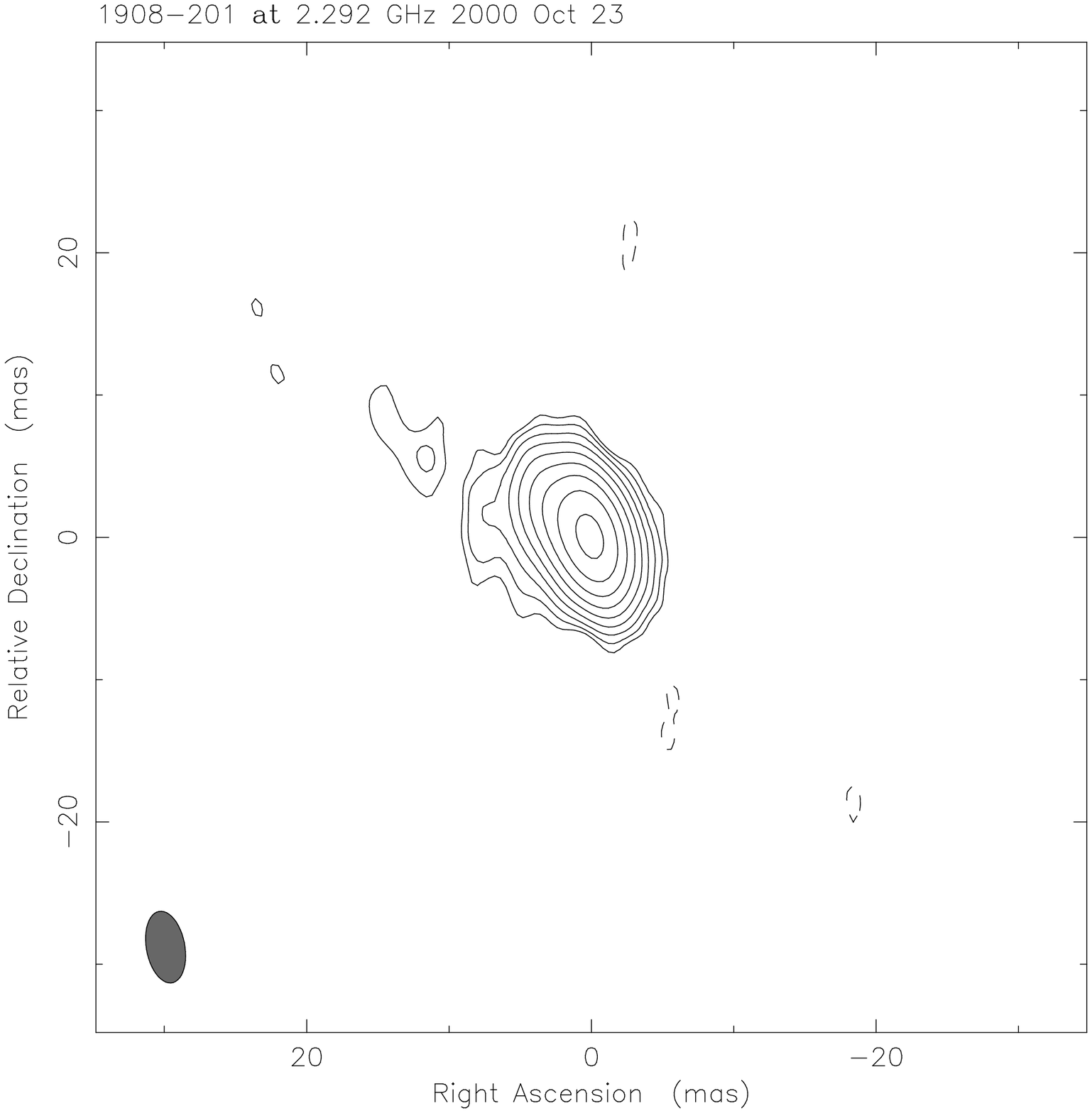}

 \includegraphics[height=7.6cm]{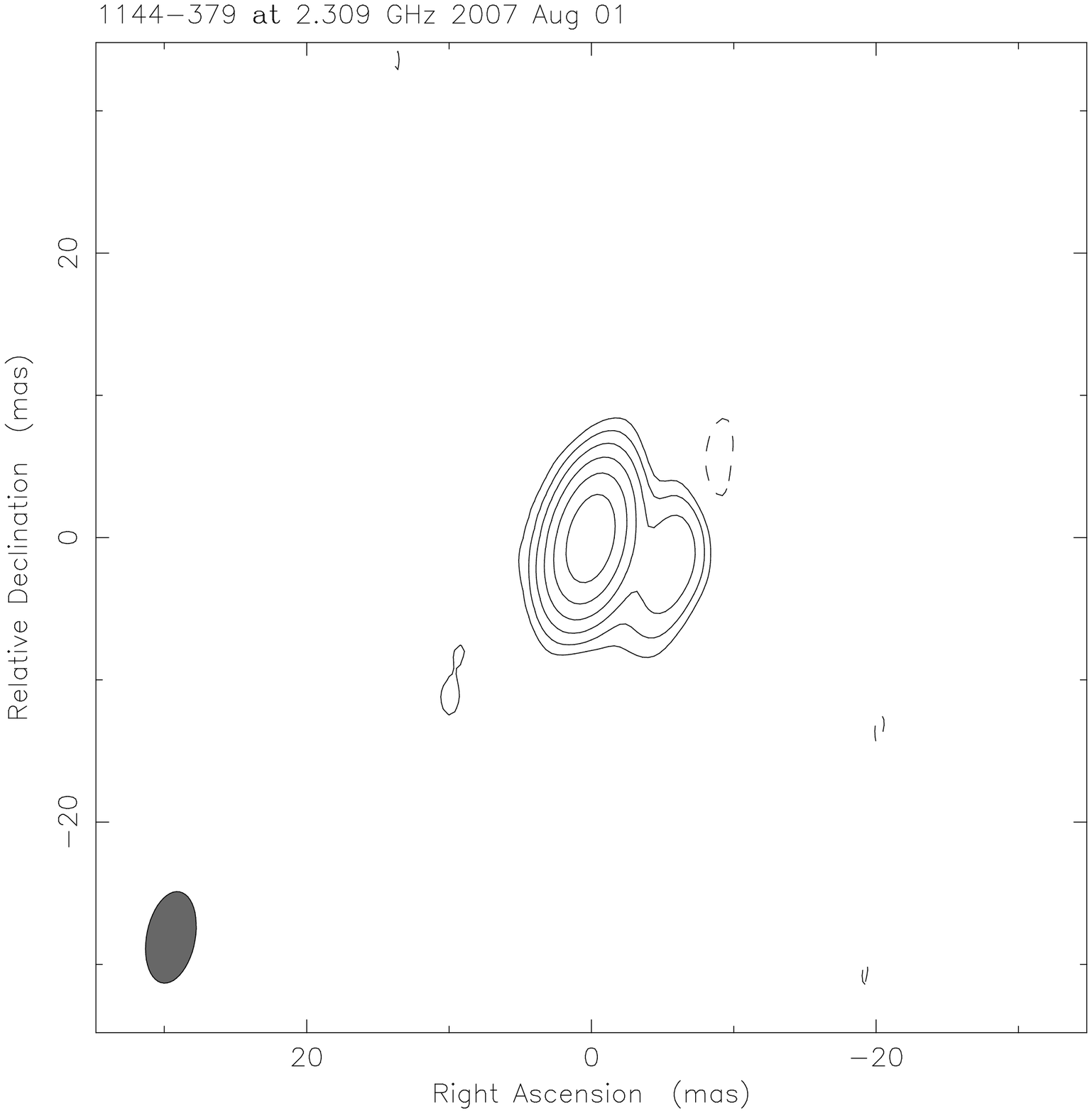}
 \includegraphics[height=7.6cm]{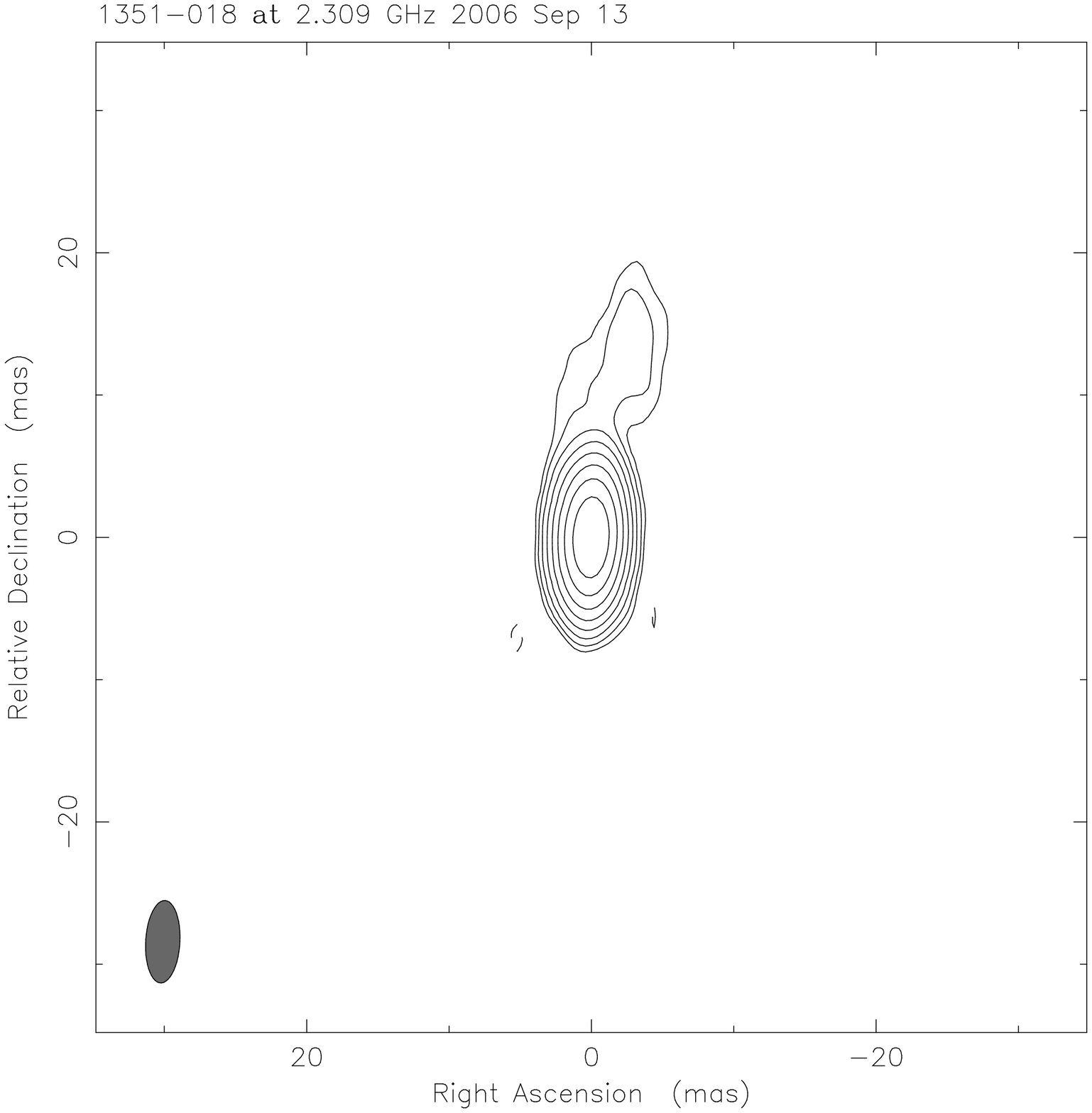}

 \includegraphics[height=7.6cm]{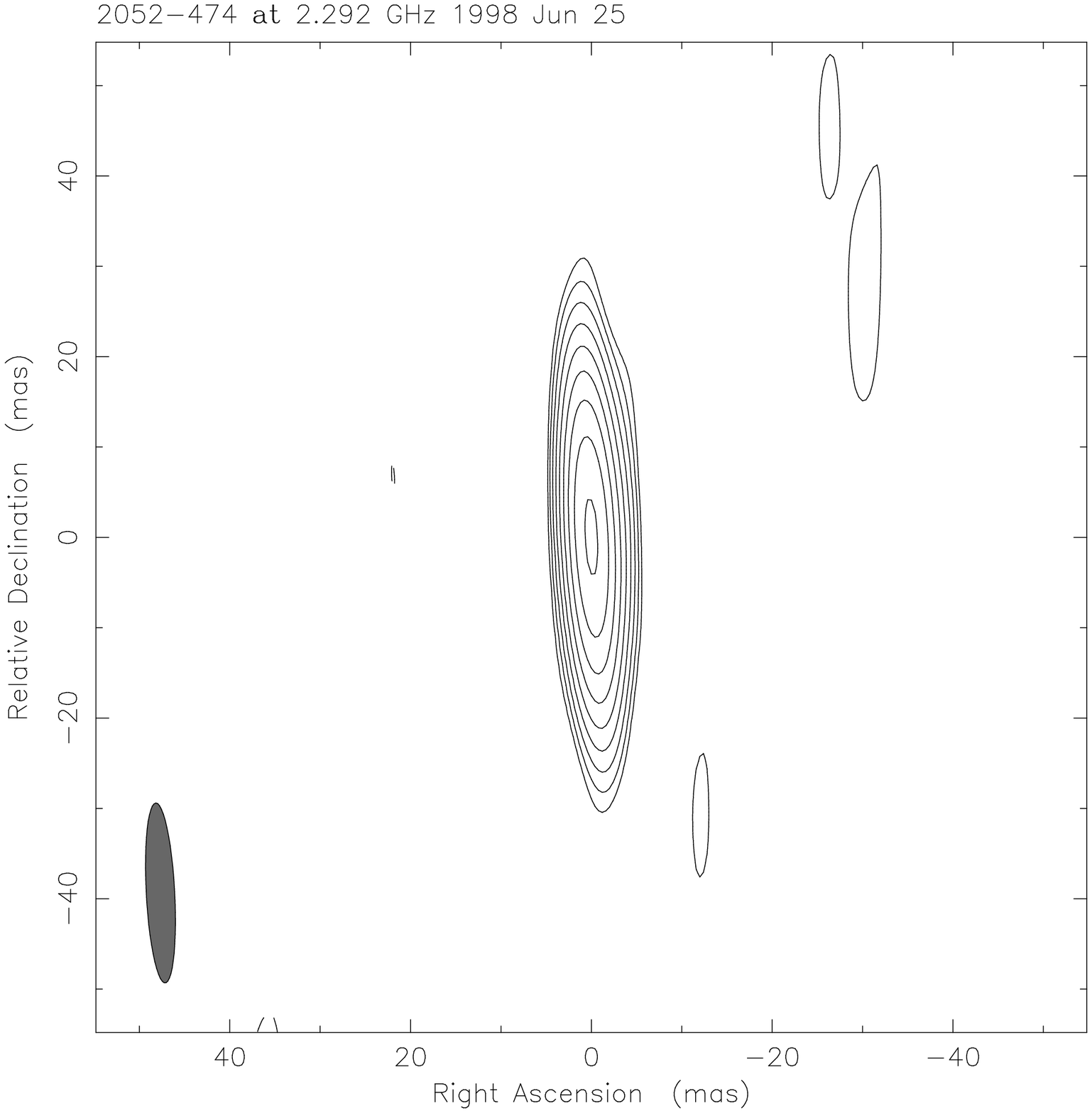}
 \includegraphics[height=7.6cm]{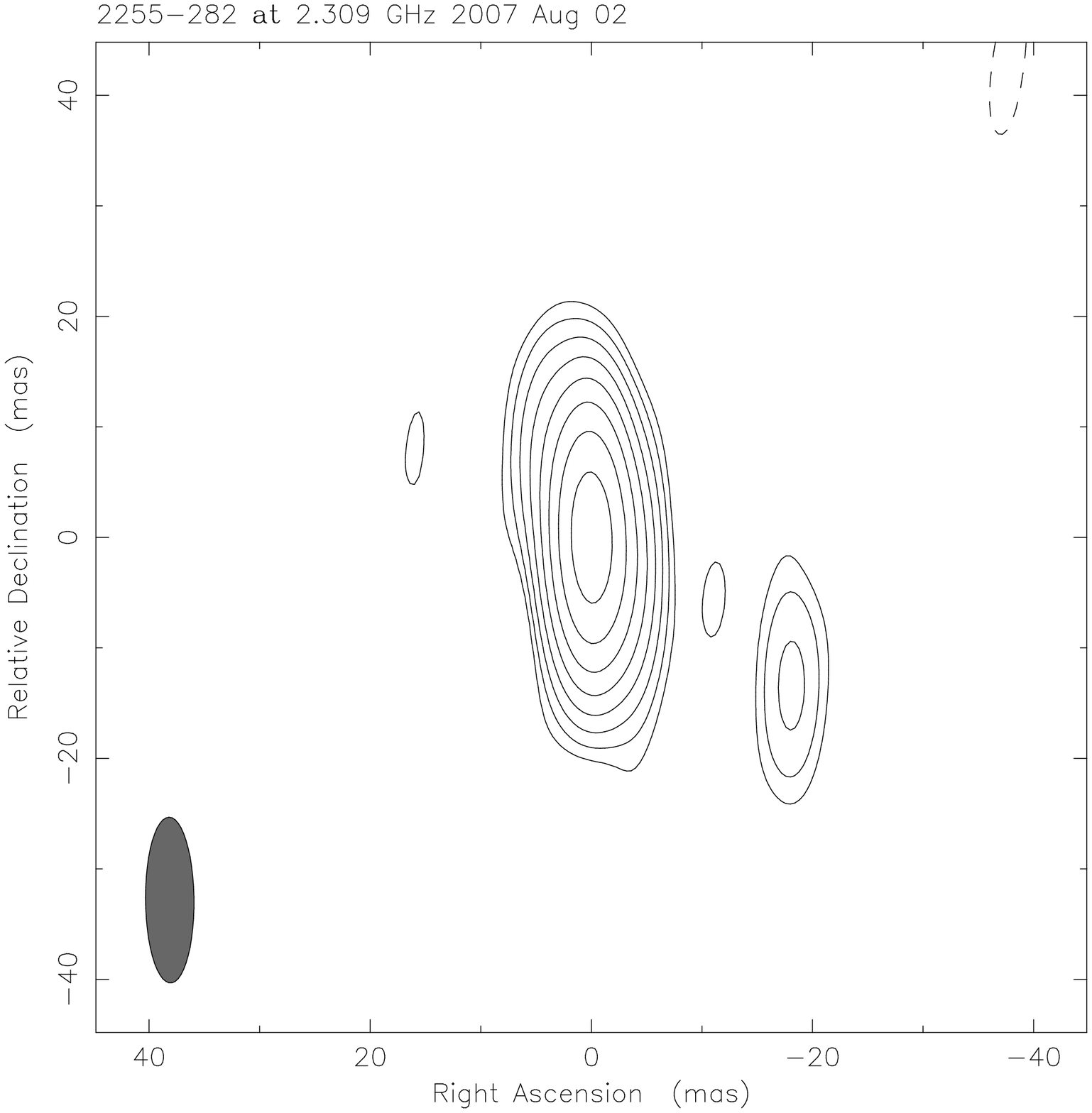}
 \caption{Two examples each of class `A', `B', and `C' sources, respectively. Our current sample contains no examples of classes `D' and `E'.}
 \label{images} 
 \end{figure*}

Research and Development VLBI (RDV) experiments are carried out using the 10 antennas of the NRAO's VLBA and up to 10 other antennas across the globe, including Hartebeesthoek (South Africa) when available. The use of the global array greatly improves the \uv\ coverage. Hartebeesthoek greatly improves the \uv\ coverage for sources in the south, which make up the sample in this paper.  

The 31 southern sources included in our sample are listed in Table~\ref{tsouthernsample} along with their optical properties. The source sky distribution is shown in Figure~\ref{skycoverage}. The RDV observing epochs are from July 1994 to January 2008, giving a total of 32 epochs and an average of 20 epochs per source. About 100 sources were observed in each 24\,hr observing run, with an average on-source time of 15~minutes. The on-source time is not continuous, but divided into scans of
between 1 and several minutes long, spaced in time to give optimal \uv\ coverage.

Observations were made in a dual frequency bandwidth synthesis mode to facilitate delay measurements for astrometry. Observations in this mode also allow imaging 
at both frequency bands. Eight individual frequency bands (IFs) were recorded simultaneously, each 8~MHz wide, with 4 at 2.3~GHz and 4 at 8.4~GHz for a total of 32~MHz in each frequency band. 

The data were correlated with the VLBA correlator at the Array Operations Center in Socorro, New Mexico. The correlated data were calibrated and corrected for time and frequency dependent phase variations using NRAO's Astronomical Imaging Processing System (AIPS, \citet{Greisen:98}). Initial amplitude calibration for each IF was accomplished using system temperature measurements and gain curves generated during observations. Fringe-fitting was done in AIPS using a solution interval equal to the scan duration and a point source model. 

Amplitude calibration was improved in a second stage by using observations of sources whose core flux density is known to be $\ge 90$ per cent of the total flux density. For a precise definition of the core, refer to Section 3.1. To this end, a single amplitude gain correction factor was derived for each antenna based on fitting a Gaussian model to the core component. Gain correction factors were then calculated based on the difference between the observed and the model visibilities. Finally, the amplitude gain correction factors were applied to the target sources. This is a non-standard procedure that improves the overall amplitude calibration. The accuracy of the amplitude calibration determined in this way is conservatively estimated to be within 20 per cent.

Data imaging and modelfitting were done using the Caltech Difference
Mapping program `Difmap' \citep{Shepherd:95}, after inspecting the data and editing out obvious bad points. 
The data were imaged using Difmap in automatic mode. Generally, this mode fails for about one third of the sources which have structure that is complex or too extended for the automatic script to handle. These have to be redone by hand in interactive mode. For the southern sources in this sample, almost all had to be imaged by hand due to the poor \uv\ coverage.

 Difmap combines the visibilities in all four 2.3~GHz receiver IFs but does not correct for spectral index effects. It was assumed that the source structure variations across the IFs were negligible (6\% variation in frequency at this band). Uniform weighting was used for the initial phase self-calibration before changing to natural weighting. For our arrays, uniform weighting gives more weighting to the longer baselines whilst natural weighting gives more weighting to the shorter baselines. Images of six of the 31 sources in our sample are shown in Figure \ref{images} with the remainder available online.
Images of additional epochs can be found in the form of contour plots at the RRFID and BVID websites.

Generally, circular Gaussian models were used to fit the \uv\ data in order to parametrise the source morphology. Like imaging, modelfitting is an iterative process. Elliptical Gaussian components were used only to represent the core component or a very bright jet component if the residuals remaining from a circular Gaussian model were too large and made it difficult to continue modelfitting using the residual map. The modelfits generally describe the visibility data well but these models may not be unique because of incomplete sampling in the \uv\ plane. In order to determine
suitability of a source as a calibrator, we then determined the amount of source structure, as well as its variation with time, using several different methods based on both the CLEAN components from the Difmap imaging as well as on the parameters of the fitted Gaussians.

\section[]{Analysis and Results}
Characterisation of the morphology of an AGN is very complex and no single metric is able to adequately define a good calibrator. Thus we have developed a number of complementary approaches that are described below. A weighted combination of all of these metrics is eventually used to classify the suitability of each source to be a high resolution phase calibrator. These are discussed below. 
\begin{table*}
\centering
\begin{tabular}{|l|c|ccc|ccc|ccc|ccc}
\hline
    Source&Epochs&\multicolumn{3}{|c|}{Core Flux Density}&\multicolumn{3}{|c|}{Core fraction}&\multicolumn{3}{|c|}{Weighted Radial Extent}&\multicolumn{3}{|c|}{Unweighted Radial Extent}\\
Name&&$\bar{S}$&$\sigma_{core}$ & $\sigma_{core}/$    & $\bar{C}$&     $\sigma_{C}$&    $\sigma_{C}/$ &	 $\bar{R}_{W}$&$\sigma_{R_{W}}$&	$\sigma_{R_{W}}/$&$\bar{R}_{UW}$ &$\sigma_{R_{UW}}$&  $\sigma_{R_{UW}}/$\\
&&&&$\bar{S}$&&&$\bar{C}$&&&$\bar{R}_{W}$&&&$\bar{R}_{UW}$\\
\hline
\hline
0003-066&24 &1.55 &0.31 &0.20 &0.64 &0.07 &0.11 &1.67 &0.40 &0.24  &6.17 &2.22 &0.36\\
0104-408&28 &1.26 &0.29 &0.23 &1.00 &0.04 &0.04 &0.93 &1.24 &1.33  &7.86 &17.13 &2.18\\
0238-084&16 &0.91 &0.20 &0.22 &0.53 &0.10 &0.19 &0.37 &0.10 &0.27  &17.34 &16.13 &0.93\\
0336-019&25 &1.92 &0.48 &0.25 &0.80 &0.08 &0.10 &1.00 &0.64 &0.64  &3.05 &2.32 &0.76\\
0402-362&17 &1.22 &0.11 &0.09 &1.00 &0.03 &0.03 &1.18 &0.26 &0.22  &8.11 &4.54 &0.56\\
0454-234&27 &1.73 &0.59 &0.34 &1.00 &0.02 &0.02 &0.46 &0.31 &0.67  &3.18 &1.78 &0.56\\
0458-020&26 &0.78 &0.31 &0.40 &0.68 &0.15 &0.22 &2.22 &1.02 &0.46  &16.30 &17.93 &1.10\\
0727-115&32 &2.37 &0.71 &0.30 &0.83 &0.05 &0.06 &1.15 &0.30 &0.26  &6.10 &1.77 &0.29\\
0919-260&17 &1.32 &0.83 &0.63 &0.73 &0.11 &0.15 &1.70 &0.34 &0.20  &6.46 &1.55 &0.24\\
0920-397&16 &1.06 &0.18 &0.17 &0.92 &0.11 &0.12 &1.81 &0.94 &0.52  &7.27 &2.98 &0.41\\
1034-293&27 &1.11 &0.30 &0.27 &0.88 &0.21 &0.24 &0.91 &0.68 &0.75  &5.83 &8.22 &1.41\\
1124-186&26 &0.90 &0.19 &0.21 &1.00 &0.02 &0.02 &0.25 &0.29 &1.18  &6.88 &16.50 &2.40\\
1144-379&23 &1.34 &0.39 &0.29 &0.83 &0.05 &0.06 &0.44 &1.79 &4.10  &9.61 &20.66 &2.15\\
1145-071&16 &0.69 &0.11 &0.16 &0.77 &0.10 &0.13 &1.33 &0.24 &0.18  &6.11 &2.26 &0.37\\
1253-055&3 &6.75 &0.27 &0.04 &0.65 &0.15 &0.23 &2.27 &0.25 &0.11  &12.5 &0.50 &0.04\\
1255-316&14 &1.20 &0.24 &0.20 &0.79 &0.11 &0.14 &2.68 &0.67 &0.25  &14.12 &11.86 &0.84\\
1313-333&17 &0.80 &0.28 &0.35 &0.71 &0.05 &0.07 &1.29 &0.40 &0.31  &7.82 &3.05 &0.39\\
1334-127&25 &2.12 &0.55 &0.26 &1.00 &0.04 &0.04 &0.84 &0.62 &0.74  &3.60 &1.87 &0.52\\
1351-018&13 &0.78 &0.25 &0.32 &1.00 &0.03 &0.03 &1.51 &2.93 &1.94  &2.55 &1.86 &0.73\\
1424-418&18 &1.69 &0.49 &0.29 &0.73 &0.08 &0.11 &3.42 &1.88 &0.55  &22.55 &11.07 &0.47\\
1451-375&14 &1.11 &0.30 &0.27 &0.88 &0.07 &0.08 &0.37 &3.71 &10.08  &9.31 &5.40 &0.58\\
1514-241&16 &1.76 &0.30 &0.17 &0.83 &0.05 &0.06 &3.39 &0.78 &0.23  &18.57 &3.90 &0.21\\
1622-253&24 &1.27 &0.38 &0.30 &0.83 &0.05 &0.06 &1.26 &0.72 &0.57  &6.35 &7.37 &1.16\\
1741-038&28 &3.18 &1.05 &0.33 &1.00 &0.03 &0.03 &0.33 &0.22 &0.66  &2.17 &1.24 &0.57\\
1908-201&23 &1.95 &0.45 &0.23 &0.77 &0.07 &0.09 &1.50 &0.45 &0.30  &5.60 &1.68 &0.30\\
1921-293&23 &7.42 &2.45 &0.33 &0.60 &0.09 &0.15 &2.30 &0.53 &0.23  &8.78 &8.17 &0.93\\
1954-388&21 &2.16 &0.54 &0.25 &1.00 &0.05 &0.05 &0.56 &0.52 &0.93  &3.52 &2.71 &0.77\\
1958-179&9 &1.16 &0.52 &0.45 &0.80 &0.04 &0.05 &0.17 &0.56 &3.29  &2.79 &2.29 &0.82\\
2052-474&10 &1.36 &0.34 &0.25 &0.81 &0.29 &0.36 &1.33 &2.09 &1.56  &10.45 &10.66 &1.02\\
2243-123&22 &1.50 &0.33 &0.22 &0.86 &0.06 &0.07 &2.00 &0.48 &0.24  &9.32 &1.77 &0.19\\
2255-282&20 &0.95 &1.11 &1.17 &0.77 &0.10 &0.13 &2.30 &2.05 &0.89  &6.58 &3.49 &0.53\\
\hline   
         
\hline            
\end{tabular}
\caption{Results of variability analysis in the flux density, core fraction and radial extent.}
\label{standarddev}
\end{table*}

\subsection{Core Flux Density}
A good calibrator should be relatively bright at the frequency of observation to be easily detectable. It should also be stable with minimal flux density variation over time. There is no precise definition of the AGN `core' in the literature. In general the bright, compact flat spectrum feature is referred to as the core. We have confirmed our identification of the cores for our entire sample by establishing that these structures have flat or inverted spectra. This was done by using the 8.4\,GHz data that are observed simultaneously with the 2.3\,GHz observations presented here. Here we define a core flux density, $S_{core}$ for each source and for each epoch as the flux density of the Gaussian component fitted to the core. The latest core flux density as defined in this section is shown, for each source in Table~\ref{tsouthernsample}.
From the core flux density, we computed the mean core flux density ($\bar{S}$) averaged over all epochs in which the source was observed. The extent to which the core flux density varies over time can be characterised by the core flux density variability index ($\sigma_{core}/\bar{S}$) where $\sigma_{core}$ is the core flux density standard deviation. A value of 0.0 indicates no variation over time (Table~\ref{standarddev}, column 5). 

\begin{figure}
 \centering
 \includegraphics[width=6cm, angle=-90]{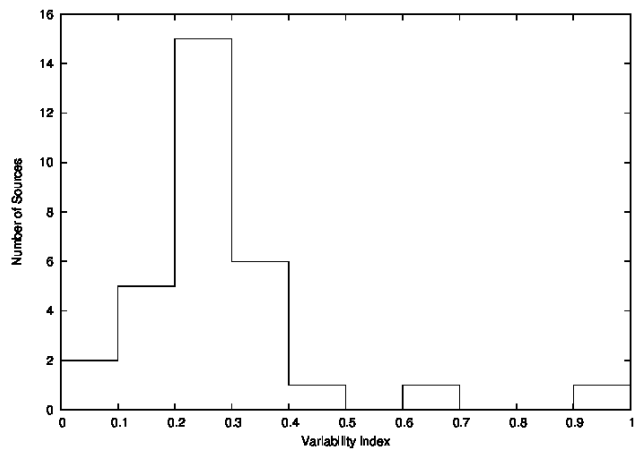}
 \caption{Distribution of the core flux density variability index.}
 \label{corehist} 
 \end{figure}
 
Figure~\ref{corehist} shows that 90 per cent of the sources have a core flux density variability index below 0.4 and 23 per cent have a variability index below 0.2 (which is below our estimated calibration uncertainities). These low variability indices indicate that most core fluxes in this sample are stable.

\subsection{Core Fraction}
Following \citet{Ojha:04b} we define a core fraction as:
 
\begin{equation}
C = S_{core\_cc}/S_{total}
\end{equation}
where $S_{core\_{cc}}$ is the sum of the flux densities of CLEAN components within one synthesised beam of the brightest pixel and $S_{total}$ is the sum of all the CLEAN component flux densities.

It provides an indication of how point-like a source is and also provides a way to track source structure changes from epoch to epoch.  The average core fraction, $\bar{C}$, was computed for each source over all epochs. As with the core flux density variability index, we also computed the source core fraction variability index ($\sigma_{C}/\bar{C}$) where $\sigma_{C}$ is the core fraction standard deviation. A value of 0.0 for the source core fraction variability index indicates no variation over time. The results are shown in Table~\ref{standarddev}, column 8.

\begin{figure}
 \centering
 \includegraphics[width=6cm, angle=-90]{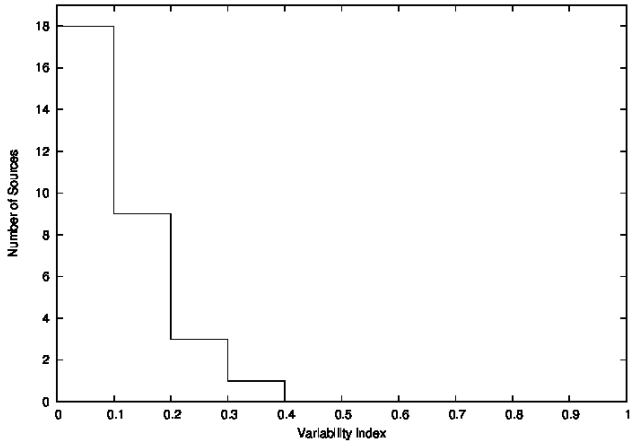}
 \caption{Distribution of the core fraction variability index. Note that all sources have an index $\le$ 0.4.}
 \label{corefrac} 
 \end{figure}
 
\begin{figure}
 \centering
\includegraphics[width=6cm,angle=-90]{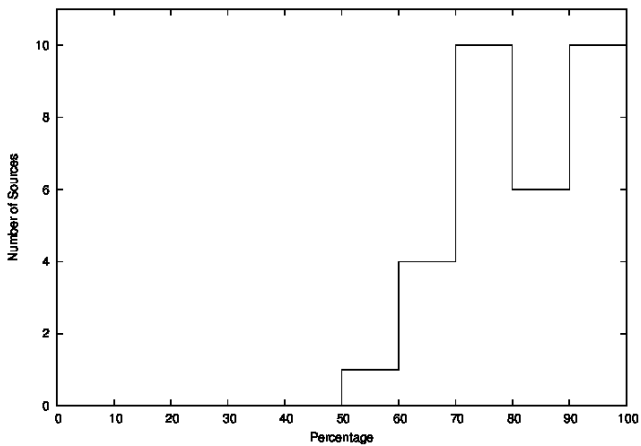}
 \caption{Distribution of the core fraction. This measure was used instead of the variability index as it shows slightly higher variation than the core fraction variability index. }
 \label{compacthist} 
 \end{figure}
The distribution of the core fraction variability index is shown in Figure~\ref{corefrac} while the distribution of the core fraction is shown in Figure \ref{compacthist}. The mean core fraction for all the sources is 83 per cent with a standard deviation of 12 per cent. All sources have a mean core fraction variability index below 0.40. Twenty seven sources have a variability index between 0 and 0.2 and the remaining 4 between 0.21 and 0.40. 
In general, the southern sample sources are very compact with little variation. For this reason, we used the actual core fraction and not the variability index to classify the sources. 

\subsection{Flux Weighted Radial Extent}
The previous two metrics are primarily measures of core-dominance. This and the following metric provide complementary information by quantifying the radial extent of the sources in the Southern sample.

The flux density weighted radial extent \citep{Ojha:04b} is defined as:
\begin{equation}
 R_{W} = \frac{\sum_{i} {S_{i} r_{i}}}{\sum_{i}S_{i}}  
\end{equation}
where $R_{W}$ is in units of milliarcseconds and $r_{i}$ is the radius at which the $i$th CLEAN component has flux density $S_{i}$.

The mean ($\bar{R}_{W}$) and standard deviation were also calculated as was the variability index ($\sigma_{R_{W}}/\bar{R}_{W}$) where $\sigma_{R_{W}}$ is the weighted radial extent standard deviation, see Table~\ref{standarddev}, column 11.

\begin{figure}
 \centering
 \includegraphics[width=6cm,angle=-90]{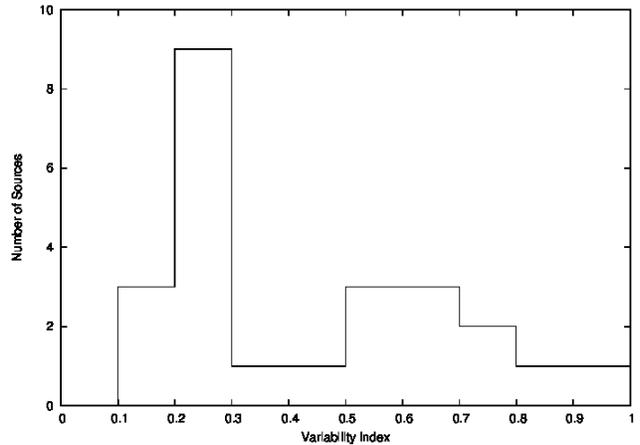}
 \caption{Distribution of the Weighted radial extent variability index. Seven sources with variability index  greater than 1 are not shown in the histogram.}
 \label{weightedhist} 
 \end{figure}
The weighted radial extent variability index is widely distributed as shown in Figure~\ref{weightedhist}, with 10 per cent of the sources having an index between 0 and 0.2, showing minimal variation in the weighted radial extent. 32 per cent of the sources have an index between 0.21 and 0.40 while 13 per cent have an index between 0.41 and 0.60. 16 per cent of the sources have an index between 0.61 and 0.80 while the rest, (29 per cent) have an index greater than 0.81. This shows a high degree of variability in the weighted radial extent.

\subsection{Unweighted Radial Extent}
This is the radial extent within which 95 per cent of the source flux density is contained. This provides a measure of how extended a source is. The mean ($\bar{R}_{UW}$) and standard deviation were calculated as was the variability index ($\sigma_{R_{UW}}/\bar{R}_{UW}$) where $\sigma_{R_{UW}}$ is the unweighted radial extent standard deviation  and $R_{UW}$ is in units of milliarcseconds, (Table~\ref{standarddev}, column 14)

\begin{figure}
 \centering
 \includegraphics[width=6cm,angle=-90]{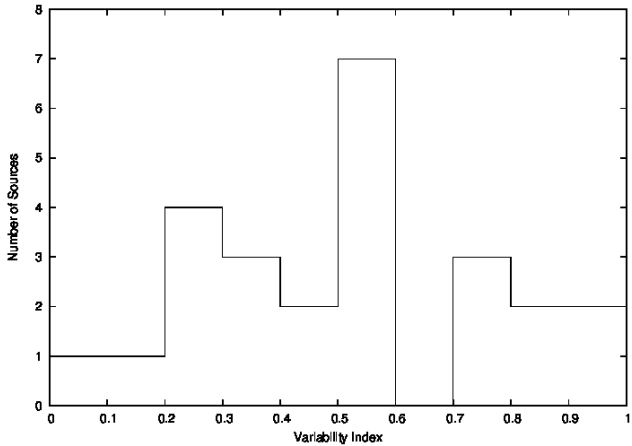}
 \caption{Distribution of the Unweighted radial extent variability index. Six sources with variability index greater than 1 are not shown in the histogram.} 
 \label{unweighthist}%
 \end{figure}
Like the weighted radial extent  variability index (see above), the unweighted radial extent variability index is also widely distributed. Figure~\ref{unweighthist} shows that 6 per cent of the sources have an index below 0.2, 23 per cent have an index between 0.21 and 0.40 while 29 per cent have an index between 0.41 and 0.60. About 10 per cent have an index between 0.61 and 0.80 while 35 per cent have an index greater than 0.80, showing very high variability in the unweighted radial extent. As discussed below, this high variability (in both the weighted and unweighted radial extent) is largely a result of low brightness `jet' features that are detected in some epochs but not others, usually due to differing sensitivities of the observing array rather than a real variation in the source. 

\section{Source Classification}
To capture the complexity of determining whether a radio source is a suitable phase calibrator for a radio interferometer, we have developed the set of complementary metrics described above. Here we discuss how we combine the information these metrics provide to arrive at a classification scheme for potential calibrators. We explain our reasons for weighting the different metrics as we have, and present our recommendations based on this weighting scheme. Since such a scheme is necessarily somewhat subjective it should be considered as reasonable and useful rather than a definitive classification. 

\begin{table}
\begin{center}
\begin{tabular}{llll}
\hline
Core flux density & Score&Core fraction &Score\\
(\%)&&variabilty Index&\\
\hline
\hline 
$>1$&0.00&&\\
0.91-1.00&3.5&91-100 &35.0 \\ 
0.81-0.90&7.0&81-90&31.5  \\ 
0.71-0.80&10.5&71-80 &28.0\\ 
0.61-0.70&14.0&61-70 &24.5 \\ 
0.51-0.60&17.5&51-60 &21.0 \\
0.41-0.50&21.0&41-50&17.5\\
0.31-0.40&24.5&31-40&14.0\\
0.21-0.30&28.0&21-30&10.5\\
0.11-0.20&31.5&11-20&7.0\\
0.01-0.10&35&1-10&3.5\\
\hline
\end{tabular}
\end{center}
\caption{Source core fraction  and core flux density variability index score distribution.}
\label{compact}
\end{table}

\begin{table}
\begin{center}
\begin{tabular}{ll}
\hline
Index & Score \\ 
\hline
\hline
$>1$&0.00\\
0.91-1.00 & 1.5 \\ 
0.81-0.90 & 3.0 \\ 
0.71-0.80 & 4.5 \\ 
0.61-0.70 & 6.0\\ 
0.51-0.60& 7.5\\
0.41-0.50&9.0\\
0.31-0.40&10.5\\
0.21-0.30&12.0\\
0.11-0.20&13.5\\
0.01-0.10&15\\
\hline
\end{tabular}
\end{center}
\caption{Distribution of scores for the weighted and unweighted radial extent variability index.}
\label{radialex}
\end{table}

The four metrics we used to classify the calibrator source quality
were core fraction, core flux density variability index, weighted and
unweighted radial extents.  Each of these metrics is first assigned a
score, with the scores as shown in Tables~\ref{compact} and~\ref{radialex}.  The scores for the
individual metrics were chosen so as to sum to 100 for a perfect
calibrator source.  We also chose to score the core fraction and core
flux density variability so as to give them higher weight, each having
a maximum score of 35, while the two radial extents have maximum
scores of 15.  We use a lower maximum score for the radial extents as
they are highly sensitive to small epoch-to-epoch variations in the
signal-to-noise ratio of the images.  Faint extended features may be
detected only at some epochs depending on the signal-to-noise,
resulting in a large variation in the radial-extent measures.  Such
low surface brightness features generally have negligible impact on
the usefulness of the source as a calibrator. In any case, for a lower resolution array like meerKAT these extended structures will be embedded in the main central component. 

Based on  the overall score summed up from this weighting scheme each source falls into one of five classes `A' through `E'. 
A source falls into class `A' if its overall score is between 80 and 100 per cent, class `B' if the score is between 60 and 80 per cent, class `C' if it scores between 40 and 60 per cent, class `D' for any score between 20 and 40 per cent and class `E' for sources scoring between 0 and 20 per cent. We propose the following classifications:


 \begin{itemize}
 \item[A] - Excellent calibrator (score between 80 and 100).
 \item[B] - Very good calibrator (score between 60 and 80 ).
 \item[C] - Good calibrator (score between 40 and 60).
 \item[D] - Use with caution (score between 20 and 40).
\item[E] - Unsuitable as calibrator (score below 20).
\end{itemize}

In our sample (Table~\ref{classification}), we have 9 class `A' sources, 19 class `B' sources and 3 class `C' sources. There are no sources in class `D' and class `E'. Thus all the sources in our sample would be `Good' calibrators and all but three sources are likely to be `Very Good' calibrators. 

\begin{table}
\begin{center}
\begin{tabular}{|l|c|c|c|c|c|c|}
\hline
Source &\multicolumn{4}{c}{Score}&Total Score&Class\\
Name& W$_{1}$    & W$_{2}$    &W$_{3}$& W$_{4}$&\%&\\
\hline
0003-066&31.5		&31.5		&10.5		&12.0	&85.5	&A	\\
0104-408&28.0		&35.0		&0.0			&0.0	        &63.0	         &B	\\
0238-084&	28.0		&31.5		&1.5			&12.0	&73.0	&B	\\
0336-019&	28.0		&35.0		&4.5			&6.0	         &73.5	&B	\\
0402-362&	35.0		&35.0		&7.5			&12.0         &89.5	&A	\\
0454-234&	24.5		&35.0		&7.5			&6.0	         &73.0	&B	\\
0458-020&24.5		&31.5		&0.0			&9.0	         &65.0	&B	\\
0727-115&	28.0		&35.0		&12.0		&12.0	&87.0	&A	\\
0919-260&14.0		&31.5		&12.0		&13.5	&71.0	&B	\\
0920-397&31.5		&31.5		&9.0			&7.5  	&79.5	&B	\\
1034-293&28.0		&28.0		&0.0			&4.5  	&60.5	&B	\\
1124-186&28.0		&35.0		&0.0			&0.0  	&63.0	&B	\\
1144-379&28.0		&35.0		&0.0			&0.0  	&63.0	&B	\\
1145-071&31.5		&31.5		&10.5		&13.5	&87.0	&A	\\
1253-055&35.0		&28.0		&15.0		&13.5	&91.5	&A	\\
1255-316&	31.5		&31.5		&3.0			&12.0 	&78.0	&B	\\
1313-333&	24.5		&35.0		&10.5		&10.5	&80.5	&A	\\
1334-127&	28.0		&35.0		&7.5			&4.5  	&75.0	&B	\\
1351-018&	24.5		&35.0		&4.5			&0.0  	&64.0	&B	\\
1424-418&	28.0		&31.5		&9.0			&7.5  	&76.0	&B	\\
1451-375&	28.0		&35.0		&7.5			&0.0  	&70.5	&B	\\
1514-241&	31.5		&35.0		&12.0		&12.0	&90.5	&A	\\
1622-253&	28.0		&35.0		&0.0			&7.5  	&70.5	&B	\\
1741-038&	24.5		&35.0		&7.5			&6.0  	&73.0	&B	\\
1908-201&	28.0		&35.0		&12.0		&12.0	&87.0	&A	\\
1921-293&	24.5		&31.5		&1.5			&12.0	&69.5	&B	\\
1954-388&	28.0		&35.0		&4.5			&1.5  	&69.0	&B	\\
1958-179&	21.0		&35.0		&3.0			&0.0  	&59.0	&C	\\
2052-474&	28.0		&24.5		&0.0			&0.0  	&52.5	&C	\\  
2243-123&28.0	         &35.0	         &13.5	         &12.0	&88.5	&A	\\
2255-282&00.0		&31.5	         &7.5  	         &3.0  	&42.0	&C	\\	
\hline
\end{tabular}
\end{center}
\caption{Classification of the Sources. W$_{1}$ - Core Flux Density,
W$_{2}$ - Core fraction,
W$_{3}$ - Unweighted Radial Extent,
W$_{4}$ - Weighted Radial Extent}
\label{classification}
\end{table}

\section{Conclusions}
 We modelfitted up to 32 epochs of observations (average of 20 epochs) for each of the 31 sources in our sample which was selected from the RRFID Kinematic Survey \citep{Piner:07} and determined their suitability as phase calibrators. While the kinematic survey looks more at proper motion in the sources at 8.4~GHz, this paper concentrates on the morphological properties of the sources at 2.3~GHz, a frequency more relevant to emerging southern hemisphere arrays like meerKAT and ASKAP. 
 
We have developed a method to classify radio sources according to
their suitability as phase calibrators for radio interferometers. We
first characterize a source by calculating several metrics which give
measures of the degree to which the core dominates the source, and the
degree of variability, both in flux density, and the degree to which
the source is extended.  These metrics are then combined to give the
source a total score, which is used to assign the source to one of
five classes of suitability as a calibrator. All 31 sources in our sample were classified as `Good' calibrators with 28 classified as `Very good' or better. 

\section*{Acknowledgments}
FH is supported by a grant from the South African SKA project and HartRAO. FH thanks the United States Naval Observatory for their hospitality during two visits in 2008 and 2010. This research has made use of the United States Naval Observatory (USNO) Radio Reference Frame Image Database (RRFID) and Laboratoire d'Astrophysique de Bordeaux (LAB) Bordeaux VLBI Image Database (BVID). This research has made use of NASA's Astrophysics Data System Bibliographic Services and the NASA/IPAC Extragalactic Database (NED) which is operated by the Jet Propulsion Laboratory, California Institute of Technology, under contract with the National Aeronautics and Space Administration.

\appendix

\bsp

\label{lastpage}

\end{document}